\documentclass[sigconf]{acmart} 

\AtBeginDocument{%
  }

\copyrightyear{2026}
\acmYear{2026}
\setcopyright{cc}
\setcctype{by}
\acmConference[CHI EA '26]{Extended Abstracts of the 2026 CHI Conference on Human Factors in Computing Systems}{April 13--17, 2026}{Barcelona, Spain}
\acmBooktitle{Extended Abstracts of the 2026 CHI Conference on Human Factors in Computing Systems (CHI EA '26), April 13--17, 2026, Barcelona, Spain}
\acmDOI{10.1145/3772363.3798401}
\acmISBN{979-8-4007-2281-3/2026/04}

\usepackage{float}
\usepackage{booktabs}
\usepackage{rotating}
\usepackage{pgfplots}
\pgfplotsset{compat=1.18}
\usepackage{multirow}
\usepackage{tikz}
\usetikzlibrary{shapes.geometric}
\usepackage{xcolor}
\usepackage{graphicx}

\begin{document}


\title[Feelings, Not Feel]{Feelings, Not Feel: Affective Audio–Visual Pseudo-Haptics in Hand-Tracked XR}

\author{Kristian Paolo David}
\affiliation{%
  \institution{University of the Philippines Los Baños}
  \city{Manila}
  \country{Philippines}}
\email{kpdavid3@up.edu.ph}

\author{Tyrone Justin {Sta Maria}} 
\orcid{0000-0001-9341-6088}
 \affiliation{%
  \institution{De La Salle University}
  \city{Manila}
  \country{Philippines}
  }
\email{tyrone_stamaria@dlsu.edu.ph}

\author{Mikkel Dominic {Gamboa}}
\orcid{0009-0004-7009-2673}
\affiliation{%
  \institution{De La Salle University}
  \city{Manila}
  \country{Philippines}
  }
\email{mikkel_gamboa@dlsu.edu.ph}

\author{Jordan Aiko {Deja}} 
\orcid{0000-0001-9341-6088}
 \affiliation{%
  \institution{De La Salle University}
  \city{Manila}
  \country{Philippines}
  }
\email{jordan.deja@dlsu.edu.ph}


\begin{abstract}
Hand-tracking enables controller-free XR interaction but does not have the tactile feedback controllers provide. Rather than treating this solely as a missing-sensation problem, we explore whether pseudo-haptic cues on an embodied virtual hand act as tactile or as affect substitutes that shape how interactions feel. We used a mixed reality prototype that keeps the contacted surface visually neutral, rendering cues on the hand with motion modulation for texture, color glow, and movement-coupled sound. In a within-subjects study ($n{=}12$), participants experienced 12 conditions (4 effects × 3 modalities: audio, visual, both) and reported subjective affect and cognitive demand. Participants rarely reported sustained tactile, thermal sensations, yet affect shifted systematically: rough-hot lowered valence increasing arousal, while smooth-cold produced calmer pleasant states. These findings suggest that pseudo-haptics in XR may be better understood as an affective feedback channel rather than a direct replacement for physical touch in controller-free systems.
\end{abstract}


\begin{teaserfigure}
  \includegraphics[width=\textwidth]{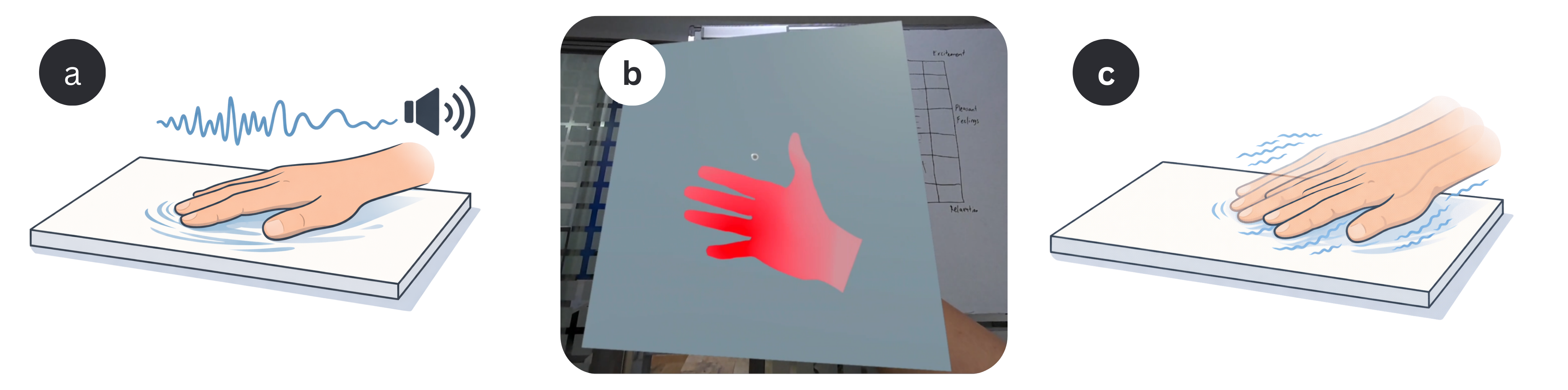}
  \caption{Hand-targeted pseudo-haptics in XR. (a) Illustration of audio cue generated by hand–surface interaction, (b) in-app screenshot thermal cues via hand color glow, and (c) illustration of tactile cues via hand motion modulation. All cues are applied to the embodied hand while the surface remains visually neutral.}
  \Description{Enjoying the baseball game from the third-base
  seats. Ichiro Suzuki preparing to bat.}
  \label{fig:teaser}
\end{teaserfigure}


\maketitle

\section{Introduction and Background}

Recent extended reality (XR) systems increasingly rely on hand tracking as a primary interaction modality, enabling direct, controller-free manipulation of virtual content~\cite{kim2025controllers}. While this shift supports more natural and accessible interaction, it also removes the incidental tactile feedback such as vibration or resistance, previously provided by handheld controllers. Although wearable haptic devices can restore physical touch sensations, they remain bulky, expensive, and impractical for everyday XR use~\cite{Xavier_Silva_Ventura_Jorge_2024}, motivating alternative feedback techniques that work within existing hardware constraints.

\par Pseudo-haptics addresses this gap by exploiting cross-modal perception, using visual and auditory cues to evoke impressions of touch without physical actuation~\cite{Lecuyer2009SurveyPseudoHaptic,Abdo_Kapralos_Collins_Dubrowski_2024}. Prior work has demonstrated that manipulating the mapping between user input and system response can convey properties such as force, weight, stiffness, and texture~\cite{Lecuyer2004BumpsHoles,Dominjon2005MassCD,Mandryk2005StickyWidgets}. However, most pseudo-haptic techniques are object-centric, modifying object behavior or appearance, and are commonly evaluated in terms of tactile substitution whether users experience sensations as if they were physically present.

\par In hand-tracked XR, pseudo-haptic cues can be anchored to the user’s embodied virtual hand rather than conveyed through a handheld controller. Strong visuomotor coupling can induce a sense of body ownership over virtual hands, drawing on mechanisms similar to the rubber hand illusion~\cite{botvinick1998rubber} and prior work on virtual limb embodiment~\cite{Yuan_Steed_2010,Argelaguet_Hoyet_Trico_Lecuyer_2016}. Under such conditions, visual and auditory cues anchored to the hand may not function purely as tactile substitutes. Instead, they may shape how interactions are affectively interpreted (e.g., intense, uncomfortable, calming, or pleasant), even when users remain aware that no physical touch is occurring.

\par Recent studies support this interpretation, showing that hand-based visual cues such as outline shape or color can systematically shift users’ affective responses without enhancing physical realism~\cite{Baeck_Shin_Kim_Lee_Yoon_Woo_2025}. These findings suggest that pseudo-haptics may operate as an affective or semantic communication channel rather than a direct replacement for physical touch. However, it remains unclear whether such affective interpretations emerge when pseudo-haptic cues are applied exclusively to the embodied hand, without any accompanying tactile feedback or object-based material manipulation.

\par In this work, we present a short empirical exploration of hand-targeted audio–visual pseudo-haptics in mixed reality. Through a controlled user study, we examine how four hand-based effects (rough, smooth, hot, cold) and three modality conditions (audio-only, visual-only, audio–visual) influence perceived valence, arousal, and attentional demand during interaction with a visually neutral surface. Our contribution is twofold: (1) initial evidence suggesting that hand-targeted pseudo-haptics can be leveraged to shape affective responses rather than inducing tactile or thermal illusions, and (2) initial design insights for using pseudo-haptics as an affective feedback layer in low-cost, hand-tracked XR systems.

\section{Study: Affective Pseudo-Haptics in XR}
\par To examine how hand-targeted pseudo-haptic cues are perceived in controller-free XR, we conducted a controlled mixed reality study. The study isolates audio and visual feedback applied exclusively to the user’s embodied hand during interaction with a visually neutral surface. This allows us to assess whether such cues are interpreted as tactile or thermal sensations or if they instead shape affect responses. 

\subsection{Hardware and Application Design}
We implemented a mixed reality prototype in Unity using the Unity XR Interaction Toolkit and deployed it on a Meta Quest~3 with optical hand tracking. Participants interacted using their tracked bare hands with a single visually neutral virtual surface (a planar square). No tactile or thermal haptic hardware were used.

\par Pseudo-haptic cues were rendered solely on the participant’s virtual hand representation; the surface of the object itself remained visually unchanged. This design isolates hand-targeted pseudo-haptics from object-centric material or behavior manipulation.

\subsection{Design and Conditions}
We employed a within-subjects design manipulating two perceptual attributes associated with touch: \textbf{tactile quality} (\textit{smooth}, \textit{rough}) and \textbf{temperature} (\textit{hot}, \textit{cold}). Each effect was presented across three sensory modality conditions: \textbf{Audio-only (A)}, \textbf{Visual-only (V)}, and \textbf{Audio-Visual (AV)}.

\par In A, only auditory cues were enabled; in V, only visual cues are enabled; and in AV, both were enabled, yielding 12 total conditions (4 effects $\times$ 3 modalities) which were counterbalanced across participants. 

\subsection{Cue Implementation}

\par \textit{Visual cues:} Tactile effects were conveyed through subtle modulation of the hand’s motion response during surface exploration. \textit{Smooth} feedback increased apparent inertia, producing slight motion overshoot during stroking gestures, while \textit{rough} feedback introduced small temporal delays and low-amplitude jitter to suggest irregular micro-contacts. Thermal effects were implemented using a colored emissive glow applied to the hand material: \textit{red} for hot and \textit{blue} for cold. No visual changes were applied to the surface.

\par \textit{Audio cues:} Audio feedback was synthesized in Unity to be suggestive of the target effect. \textit{Smooth} stimuli used continuous harmonic tones, while \textit{rough} stimuli employed noisier, granular textures. Thermal effects used distinct sound designs aligned with common hot and cold interpretations. All audio cues were temporally coupled with hand motion to preserve perceived action–feedback contingency.

\subsection{Protocol}
Twelve participants completed the study, experiencing all conditions. Conditions were presented in modality-based blocks (A, V, AV), with all four effects presented within each block. The order of modality blocks and effects within the block were independently randomized to mitigate order and carryover effects.

\par For each condition, participants explored the surface using their tracked hand (e.g., stroking with the palm or fingers) while the corresponding cues were applied. After each condition, participants reported their momentary valence and arousal using a 7$\times$7 Affect Grid.

\par After each modality block, participants completed a modified Multiple Resource Questionnaire (MReQ) to capture perceived mental demand. The questionnaire was adapted to compare relative resource allocation across modalities rather than overall workload~\cite{landgrebe2025augmenting}. Participants rated six resource channels: auditory affective processing, manual effort, visual–spatial attention, visual–temporal monitoring, spatial positioning, and short-term memory, all on a 5-point scale.

\par Following all conditions, participants took part in a brief semi-structured interview focusing on how they interpreted the cues, particularly whether they were experienced as physical touch sensations or as affective signals.

\subsection{Data Analysis}
Our analysis examined (1) whether participants reported illusionary tactile or thermal sensations, (2) how affective responses (valence and arousal) varied across effects and cue modalities, and (3) whether combined audio--visual cues increased perceived integration demand compared to single-modality cues.

\par Because the study did not include a no-cue baseline, results are interpreted as relative differences between cue types and modalities rather than absolute shifts from a neutral state. Affect Grid ratings and MReQ subscales were compared across within-subject conditions using repeated-measures ANOVA. Interview transcripts were analyzed qualitatively to distinguish reports of perceived physical sensation (e.g., rough, warm) from affective interpretation (e.g., calming, intense) and to identify recurring themes.

\section{Results and Discussion}

\definecolor{effRough}{HTML}{CC79A7}
\definecolor{effSmooth}{HTML}{009E73}
\definecolor{effHot}{HTML}{D55E00}
\definecolor{effCold}{HTML}{0072B2}

\begin{figure*} [t]
  \centering
      \includegraphics[
    width=\linewidth,
    trim=0 1.75cm 0 1.75cm,
    clip
  ]{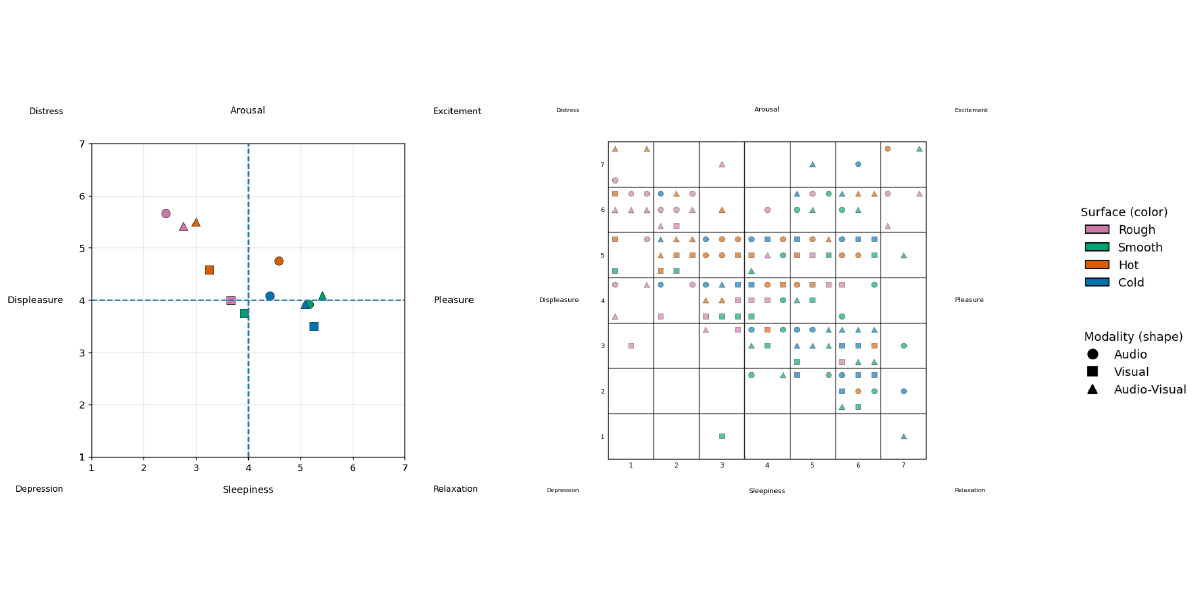}
  \caption{Affect Grid results for hand-targeted pseudo-haptic cues. Left: Mean valence–arousal ratings for each effect–modality combination. Right: Distribution of individual participant ratings, showing clustering patterns across the valence–arousal space. Color denotes effect type (rough, smooth, hot, cold), and shape indicates modality (audio, visual, audio–visual).}
  \label{fig:affectgrid_scatter_simple}
\end{figure*}

\subsection{Valence--Arousal (Affect Grid)}

Figure~\ref{fig:affectgrid_scatter_simple} shows that hand-targeted pseudo-haptic cues produced systematic shifts in affect, even when participants did not report sustained tactile or thermal sensations. Rather than clustering randomly, ratings occupied distinct regions of the valence--arousal space, ranging from low-valence/high-arousal states (e.g., tense, unpleasant) to higher-valence/low-arousal states (e.g., calm, pleasant), supporting an interpretation of these cues as primarily affective rather than sustained tactile/thermal feeling.

\par Across conditions, effect type largely determined affect direction: rough and hot cues shifted ratings toward lower valence and higher arousal, whereas smooth and cold cues clustered in higher-valence, lower-arousal regions. Modality influenced how these affective shifts manifested along each axis. Audio-inclusive conditions (Audio-only and Audio--Visual) generally produced higher arousal than Visual-only conditions, indicating that sound primarily drove emotional intensity. Visual cues, in contrast, primarily biased valence in temperature effects: blue (cold) shifted responses toward calmer, more pleasant states, while red (hot) shifted them toward lower valence. Motion-based visual hand modulation showed weaker and less consistent effects, likely because the low-amplitude jitters and subtle overshoot or delay were embedded in natural movement and therefore less salient as affective cues.

\par Importantly, combining modalities did not uniformly amplify affect. Instead, audio–visual cues interacted with the effect’s meaning: for cold and rough, using combined cues shifted responses toward a more “regulated” profile (higher valence and/or lower arousal), whereas for hot the audio–visual condition pushed ratings toward higher arousal and lower valence than either single-modality cue. Smooth showed a modest increase in both arousal and valence under audio–visual cues. Overall, these results suggest that multimodal hand-targeted pseudo-haptics along separable dimensions, and their combined effect depends on the intended affective meaning and perceived cue alignment.
\subsection{Cognitive Resource Demand (Modified MReQ)}



\begin{table*}[h]
\centering
\small
\setlength{\tabcolsep}{6pt}
\caption{Modified MReQ ratings (1--5) by resource channel and modality ($M\pm SD$). Post-hoc comparisons show direction; \textsuperscript{*}$p<.05$.}

\label{tab:mreq_means_posthoc_dir}
\begin{tabular}{lccc|ccc}
\toprule
\textbf{Channel} & \textbf{A} & \textbf{AV} & \textbf{V} & \textbf{A vs.\ AV} & \textbf{A vs.\ V} & \textbf{AV vs.\ V} \\
\midrule
Sound
& 4.17$\pm$0.83 & 4.00$\pm$0.95 & 1.75$\pm$1.22
& A $>$ AV ($p{=}0.843$)
& A $>$ V ($p{=}0.003$)\textsuperscript{*}
& AV $>$ V ($p{=}0.002$)\textsuperscript{*} \\

Move
& 4.08$\pm$0.79 & 4.08$\pm$0.79 & 4.17$\pm$0.72
& A $=$ AV ($p{=}1.000$)
& V $>$ A ($p{=}0.843$)
& V $>$ AV ($p{=}0.592$) \\

Visual
& 3.42$\pm$1.16 & 3.08$\pm$1.31 & 4.08$\pm$0.90
& A $>$ AV ($p{=}0.493$)
& V $>$ A ($p{=}0.091$)
& V $>$ AV ($p{=}0.027$)\textsuperscript{*} \\

Timing
& 3.08$\pm$1.00 & 3.58$\pm$1.24 & 4.00$\pm$0.85
& AV $>$ A ($p{=}0.289$)
& V $>$ A ($p{=}0.005$)\textsuperscript{*}
& V $>$ AV ($p{=}0.497$) \\

Control
& 3.58$\pm$1.08 & 3.67$\pm$0.78 & 3.75$\pm$1.06
& AV $>$ A ($p{=}0.930$)
& V $>$ A ($p{=}0.880$)
& V $>$ AV ($p{=}0.930$) \\

Memory
& 2.75$\pm$0.97 & 2.58$\pm$1.08 & 2.92$\pm$1.24
& A $>$ AV ($p{=}0.864$)
& V $>$ A ($p{=}0.843$)
& V $>$ AV ($p{=}0.629$) \\
\bottomrule
\end{tabular}

\vspace{2pt}
\footnotesize{\textbf{Sound:} interpret emotional qualities in sound;
\textbf{Move:} hand/arm movement;
\textbf{Visual:} attention to hand--object interaction;
\textbf{Timing:} noticing timing/animation changes;
\textbf{Control:} precise hand positioning;
\textbf{Memory:} remembering sensations from prior trials.}
\end{table*}

Table~\ref{tab:mreq_means_posthoc_dir} suggests that modality shifted where attention was allocated, rather than increasing overall perceived effort. Repeated-measures ANOVAs showed reliable modality effects for Sound ($F(2,22)=17.8, p<.001$*), Visual ($F(2,22)=5.81, p=0.009$*), and Timing ($F(2,22)=4.53, p=0.023$*), but not for Move, Control, or Memory (all $p \ge 0.599$). Post-hoc tests indicate that interpreting emotional qualities in sound was much higher when audio was present (A $>$ V, $p{=}0.003$*; AV $>$ V, $p{=}0.002$*), with high Sound ratings in Audio and AV (A: $M{=}4.17, SD{=}0.83$; AV: $M{=}4.00, SD{=}0.95$) compared to Visual-only (V: $M{=}1.75, SD{=}1.22$). In contrast, Visual-only increased visual monitoring demands: visual attention to hand–surface interaction was higher in V than AV (V $>$ AV, $p{=}0.027$*; V: $M{=}4.08, SD{=}0.90$ vs.\ AV: $M{=}3.08, SD{=}1.31$), and reliance on noticing timing/animation changes was higher in V than A (V $>$ A, $p{=}0.005$*; V: $M{=}4.00, SD{=}0.85$ vs.\ A: $M{=}3.08, SD{=}1.00$). Together, these results indicate that audio shifts workload toward sound interpretation, whereas visual-only feedback shifts workload toward sustained visual and temporal monitoring, without significantly increasing movement, positional control, or memory demands.

These resource-channel patterns are consistent with the valence-arousal modulation observed with the Affect Grid. In the modified MReQ, the Audio and Audio–Visual modalities showed higher Sound demand than Visual-only, and these audio-inclusive conditions also tended to correspond to higher arousal ratings. Visual-only feedback increased Visual and Timing monitoring demands and was associated with comparatively lower arousal overall. Together, this pattern helps explain why Audio–Visual feedback did not uniformly strengthen affect: adding audio can shift workload toward sound interpretation while reducing visual vigilance relative to Visual-only, and the resulting affect depends on the specific effect and the perceived alignment of the cues rather than simple additivity.

\subsection{Interview Insights}

Interview responses converged on a clear distinction between physical sensation and affective interpretation. Most participants explicitly stated that they did not experience sustained tactile or thermal sensations (e.g., ``I know I’m not feeling anything physically''), even when they reported strong emotional reactions. Several described brief initial impressions such as tension, flinching, or a momentary ``felt something'' followed by rapid awareness that no physical stimulus was present. This pattern aligns with our quantitative results: the cues did not reliably induce touch illusions, but they did shape how interaction felt.

\par Participants described stable cue--affect mappings consistent with the Affect Grid. Color on the hand anchored temperature as an affective association rather than a literal sensation: red was described as dangerous, burning, or stressful, while blue was calming or refreshing. Audio was most often cited as the driver of intensity, with rough or noisy sounds described as irritating or overwhelming, and smoother sounds as calming or surface-like.

\par Crucially, audio--visual outcomes depended on cue alignment. Congruent pairings were described as coherent and intuitive, while mismatches were experienced as ``weird'' or cognitively taxing. This explains why AV conditions did not uniformly amplify affect and why they exhibited higher variability in Sound, Visual, and Timing MReQ ratings. Overall, the interviews support our interpretation that hand-targeted cues primarily shaped experience through affective and associative meaning, rather than producing sustained tactile or thermal sensations.

\section{Design Implications for Affective Pseudo-Haptics}

\par Our findings suggest several implications for designing hand-targeted pseudo-haptics in controller-free XR, which align with and extend prior work on affective and multimodal feedback.

\par \textbf{Use audio as the primary driver of arousal.} Across effects and modalities, audio cues consistently increased emotional activation, shaping arousal without reliably increasing perceived workload. This aligns with prior findings showing that sound is a strong driver of affective intensity in interactive systems~\cite{warp2022moved}. Designers can therefore leverage audio to modulate arousal while keeping visual feedback lightweight.

\par \textbf{Use color-based visuals to bias valence and interpretation.} Visual cues particularly red and blue color applied to the hand effectively anchored affective interpretation of thermal qualities while exerting relatively little influence on arousal. This supports prior work showing that color primarily biases emotional meaning rather than intensity~\cite{bekhtereva2017bringing}, and suggests that simple visual properties can communicate interaction qualities without increasing perceptual load.

\par \textbf{Treat motion-based hand effects as secondary affective cues.} Subtle motion modulations on the virtual hand exhibited weaker and less consistent affective influence, and were often perceived as background interaction dynamics. This is consistent with prior observations that motion cues alone may be insufficient for affective modulation without additional salient signals~\cite{toet2022guidelines}. Designers should therefore avoid relying on motion cues in isolation, and instead combine them with audio or color-based feedback when affective impact is desired.

\par \textbf{Design multimodal feedback as adaptive rather than additive.} Audio--visual combinations did not uniformly intensify affect and showed user-dependent variability. When cues were perceived as congruent, AV felt coherent; when mismatched, it increased interpretive effort. This mirrors prior work on multimodal interaction, which emphasizes cue alignment and coherence over simple modality accumulation~\cite{schonauer2012multimodal, kim2025controllers}. Designers should treat multimodal feedback as adaptive, selectively combining cues rather than assuming that additional modalities will always improve experience.

\section{Limitations}

This study is an early exploration of hand-targeted pseudo-haptics and has several limitations. First, the sample size was modest ($n = 12$). Second, no neutral (no-cue) baseline was included; therefore, we cannot estimate absolute departure from a neutral hand–surface interaction, and results should be interpreted as relative differences across cue types and modalities rather than absolute perceptual shifts. Third, all pseudo-haptic cues were representational rather than physically grounded, limiting claims about physical realism. Fourth, the interaction context was intentionally constrained to a single visually neutral surface and hand-interaction style. While this design improves experimental control, it limits generalizability to more complex objects, tasks, and longer-term interactions. Fifth, outcomes were mainly self-report (Affect Grid, MReQ) plus interviews; we did not measure behavioral performance or objective measures such as physiological signals (e.g. galvanic skin response (GSR). Finally, individual differences in embodiment strength and sensory interpretation were not explicitly measured.

\section{Conclusion and Future Work}

\par This work explored whether pseudo-haptic cues applied to an embodied virtual hand in mixed reality induce tactile/thermal sensations or instead shape affective experience. Across four hand-targeted effects and three sensory modalities, participants rarely reported sustained touch or temperature sensations. Instead, responses consistently organized along valence–arousal dimensions, and interviews framed the experience as emotional rather than physical.

\par These findings suggest that hand-targeted pseudo-haptics in controller-free XR can function as an affective communication channel rather than a substitute for physical touch. Audio cues primarily influenced arousal, visual cues biased valence, while multimodal cues did not behave additively: outcomes varied by effect and were shaped by perceived cue alignment, with mismatches increasing effort or discomfort and aligned cues feeling more coherent.

\par Future work will examine how hand-targeted affective pseudo-haptics impacts performance and decision-making in functional tasks, expand the cue vocabulary beyond the four effects, and develop adaptive cue-selection strategies that account for cue congruence and user variability. Specifically, future study designs will incorporate a no-cue baseline to isolate the specific affective contributions of these effects against a neutral interaction state. As the present study focused primarily on subjective affective responses and perceived workload, subsequent work will incorporate objective behavioral measures to evaluate how these cues influence task performance and interaction strategies. Together, this line of work aims to provide practical design guidance for affective pseudo-haptic feedback in controller-free XR.


\bibliographystyle{ACM-Reference-Format}
\bibliography{main}

\appendix









\end{document}